\documentstyle[12pt]{article}

\begin{document}
\begin{flushright}
SU-4240-561\\
November 1993\\
\end{flushright}

\hoffset = -1truecm
\voffset = -2truecm

\vskip 2cm
\centerline{{\bf HEAVY DIQUARK EFFECTIVE THEORY AND}}

\centerline{{\bf SUPERSYMMETRY OF HADRONS CONTAINING A}}

\centerline{{\bf SINGLE HEAVY QUARK}}

\vskip 1cm

\centerline{ Nguyen Ai Viet }

\centerline{{\normalsize Department of Physics, Syracuse University, Syracuse,
New York 13244, USA}}
\vskip 3cm

\begin{abstract}
A new supersymmetry is proposed for hadrons containing a single heavy quark.
This supersymmetry is based on a new approximation to those hadrons, which we
would consider as a further step beyond the spectator light diquark model of
baryons. The heavy diquark effective theory is constructed by the techniques
introduced in a different context by Georgi and Wise$^1$ and by Carone$^2$.
This theory can be incorporated into a supersymmetric theory together with
Heavy Quark Effective Theory, and leads to a common universal Isgur-Wise
function for mesons and baryons./.
\end{abstract}

\newpage


Recently there is a considerable amount of interest in studying hadrons
containing one heavy quark ( herealfter we shall refer as 1HQ-hadrons). These
hadrons turned out to have new symmetries, which are not manifest in QCD
$^{3,4}$. So their properties are constrained to a simpler description,
the framework of the heavy quark effective theory $^{5-11}$ (HQET), where
the mass of the heavy quark $M_Q$ is taken to infinity, with its velocity held
fixed. In HQET, the heavy quark represents the kinematics of 1HQ-hadrons and is
integrated out of dynamics as a very strong static color source or Wilson line,
while the light degrees of freedom carry all the dynamics inside those hadrons.
So the dynamics of 1HQ hadrons is independent of the mass and spin of
the heavy quark. This leads to a $ SU(2N_f)$ symmetry for each heavy quark
velocity ( $N_f$ is the number of heavy flavors). Hence the long wavelength of
light degrees of freedom in such hadrons are related by this symmetry, which
gives relations between the weak form factors. The relations express the weak
form factors in term of a single reduced form factors, sometimes called the
Isgur-Wise function which maybe crucial in a
reliable extraction of the Cabbibo-Kobayashi-Maskawa matrix elements from
experimental data.\\
To push the program further to actually calculate the form factors, it is
necessary to include dynamics. Even at this stage, some more dynamical
symmetries are useful to simplify the problem to some calculable models.
Lichtenberg$^{12}$ has proposed some supersymmetries of hadrons, based on the
existence of light diquark in 1HQ-baryons. These
supersymmetries has
simplified the three-body problem of baryons to the two-body problem of mesons.
(Hereafter we shall refer to this type of supersymmetry as LSUSY).
 In this paper we shall pay special attention to
an alternative one: the heavy supersymmetry HSUSY.\\
The calculations of heavy baryons are simplified greatly by the use of the
quark-diquark model of baryons ( see$^{13}$ for the latest review on diquark).
This model has turned out to work rather well for exclusive reactions$^{14-16}$
using the fact that heavy baryons can be assumed to consist of a heavy quark
and
a light diquark. During the weak decays of the heavy quark the light diquark is
assumed to be a spectator. In other words, the light degrees of freedom are
assumed to be integrated into a pointlike diquark. So 1HQ-baryons are treated
as
a two-body system like mesons. More simplifications can be achieved by
observing
that since the light diquark and the antiquark in 1HQ-baryons and in their
meson
superpartners have the same color index, the spin and velocity-dependent
corrections to the dynamics are negligible beside the very heavy strong static
color force from the heavy quark. The calculations of Hussain, K\"orner and
Migneron $^{17}$ proved that the decay rates of baryons and mesons are equal
as a consequence of the spectator model of light degrees of freedom. This has
given an indication of the existence of supersymmetry in the model they used.
Namely, if the antiquark was replaced by a diquark with the same colour the
dynamics inside hadrons will
remain unchanged. This supersymmetry between light diquark and antiquark leads
to the LSUSY
between mesons and baryons. For light hadrons LSUSY should be broken badly
$^{18-21}$ as the differences between diquark and antiquark are not negligible
at that scale. So LSUSY is proposed to be applied for 1HQ-hadrons $^{12}$.
Here the mass of light quarks is to be compared to the mass of the heavy quark.
In such a meaning this supersymmetry would be very good ( comparing with the
usual chiral symmetry where the quark masses are compared to the scale of QCD
$\Lambda _{QCD} << M_Q$). However there are some arguments against LSUSY, not
on the diquark-antiquark supersymmetry but on the existence of the light
diquark itself:

i) The existence of a light diquark in 1HQ-hadrons is doubtful. The gluons that
the heavy quark exchanges with light degrees of freedom are soft only for the
heavy
quark but are extremely hard for the light degrees of freedom ( as their
momenta square are of
order $M_Q.\Lambda _{QCD}$). Suppose that at some moment there exists some
pointlike light diquark. The dynamics also contains hard gluon exchanges which
have enough energy to break the binding of the light diquark. If we decide to
integrate hard gluons into the diquark to have a strong binding of diquark and
a weak interaction regime with the heavy quark, we should have a very large
diquark and therefore could not approximate it by a local field.

ii) The situation is even worse for a vector diquark. As the heavy quark could
hardly rotate around the vector diquark, the centrifugal force would destroy
the
rotating vector diquark. So the light diquark picture is a poorer approximation
for $\Sigma $ and $\Xi $ baryons than it is for $\Lambda $ ones\\
iii) The calculations of Fleck et. al.$^{22}$ have indicated that in the
baryons
containing a single heavy quark the correlation of $qq$ is always much weaker
than the one of $ Qq$. It has been also shown that the heavier the baryon is,
the better is the heavy diquark approximation.

iv) Analogously, in atomic physics the $He$ atom can be well approximated as a
system of an electron and a $He^+$ ion. It can hardly be treated as a $He^{++}$
ion and a bound pair of two electrons.

Fortunately, if it is so, the heavy diquark picture will be a good
approximation
to 1HQ-baryons and it will provide a new simplification: the heavy
supersymmetry
(HSUSY). This supersymmetry in some sense is closer to the heavy quark
symmetries. The antidiquark ${\bar D}_{a,i} = ({\bar Q}{\bar q}_a)_i$ and the
quark $ Q_i$ have the same color i ( the flavour index $a$ denotes the light
flavours u,d,s) so
represent the same static color source. As the spin of this color source is
irrelevant by spin decoupling the dynamics of the remaining light degrees of
freedom
in 1HQ-hadrons should be the same and perfectly supersymmetric. This picture is
an approximation beyond the spectator model of baryons. One light quark does
not
want to be spectator anymore and approaches closer to the heavy quark to
form a heavy diquark. This heavy diquark has integrated hard gluons into
itself to build a very strong binding. So the heavy diquark is extremely stable
while the rest light quark remains spectator. As inside the diquark the spin
orientation of the heavy quark is irrelevant there is a good symmetry between
scalar and vector diquarks. As we shall see below, it is possible
to demonstrate the spin decoupling of diquarks in an effective
heavy diquark theory. It is possible to derive the low-energy
QCD, which contains the diquark fields as collective
coordinates. Such a low energy QCD can be derived from the
first principles of QCD( For instance as in the works done by Cahill$^{23}$
and Reinhardt$^{24}$ who developed an original idea of Kleinert$^{25}$ on
bilocal operators, which have been interpreted as diquarks in some
approximation). Here we shall
take only the lowest terms. Namely it is the minimal $SU(3)_{colour}$ theory of
a
scalar
field and a vector field coupled to the gluon field through the usual
Yukawa-coupling in the covariant derivative. We shall see that
in the infinity mass limit this theory will also lead to an
effective diquark theory, which is perfectly parallel with HQET
in a supersymmetric way. Therefore we can put together scalar and vector
antidiquarks into a 7-dimensional supermultiplet together with the quark. This
supermultiplet forms a spinor representation of the supergroup SU(1/6):
$$   S^i= \left (\matrix {
                           Q^i \cr
                        {\bar D}_{0a}^i \cr
                        {\bar D}_{3a}^i \cr
                            }\right )
\eqno(1)$$
where $D_{0a}^i$ and $D_{3a}^i$ denote scalar and vector diquarks with the
color
$i$ and the light flavor $a$. For the moment we shall omit the spin indices.\\

The supergroup SU(1/6) has the following decomposition:
$$  SU(1/6) \supset  U(1)\times SU(3)_{D_0}\times SU(3)_{D_3}  \eqno(2)$$
The two SU(3)-groups given in the above decomposition mix flavors of
antidiquarks. We shall differentiate between them by an index $A$, where $ A=
D_0,
D_3$. The $SU(3)_A$ group is generated by GellMann-matrices $ \lambda _{A,p} ,
p=1,...,8 $. The columm and row indices of $\lambda $-matrices will be denoted
by $\dot i,\dot j,\dot k$ and $i,j,k$ respectively.
 In the $7\times 7$ representation of the graded Lie algebra
$SU(1/6)$ we put the elements of the Lie algebras $U(1)$ and the two $SU(3)$
into
boxes lying on the diagonal of the $7\times 7$ matrix as follows:
$$  B = \left (\matrix {
 U(1)     & 0                           &0                     \cr
 0        & (SU(3)_{D_0})_i^{\dot k}         &0                     \cr
 0        &0                            & (SU(3)_{D_3})_i^{\dot k}  \cr
         }\right )
        \eqno(3)$$
The general element of the GLA $SU(1/6)$ has the following form:
$$   G = \left (\matrix {
     b_1        &f_1^+      &f_2^+  \cr
     f_1        &b_2        &b_4^+  \cr
     f_2        &b_4        &b_3    \cr
}\right )
\eqno(4)$$
where $b_1$ is a number, $b_2$ and $b_3$ are SU(3)-matrices, $f_1, f_2$ and
$b_4$
are arbitrary matrices. Beside the generators given in the equation (3) the
Bose
sector of the GLA $SU(1/6)$ also contains the generators which mix the flavors
of the scalar and vector antidiquark. We have denoted these matrices by
$(b_4)_i^{\dot k}$ and by $(b_4^+)_{\dot k}^i$. These matrices are spanned by
Gell Mann
matrices $\lambda_{4a} , (a=1,...8) $ and the $3\times 3$ unit matrix
$\lambda _4,0$. We have 36 even generators of the Bose sector. \\
The Fermi sector contains 12 odd generators: $ (f_{1,2})^{\dot k}$ and
$(f_{1,2})_k$ mixing the quark with antidiquarks. Here we ignore the spin
orientations of the quark and the diquarks. The heavy quark spin symmetry can
be incorporated in a larger supersymmetry like $SU(2/12)$.
 \\ We shall denote the 48 generators defined above by $(\beta _I)_P^Q $.
Sometimes the hermitian generators can be used instead of these non-hermitian
ones and are defined as follows:
$$ \begin{array}{lll}
              b^1_4      & =& b_4 +  b_4^+   \\
              b^2_4      & =& i ( b_4 -  b_4^+) \\
              f^1_{1,2}  & =& f_{1,2} + f_{1,2}^+ \\
              f^2_{1,2}  & =& i ( f_{1,2} - f_{1,2}^+ )  \\
              \end{array}        \eqno(5)$$
The group elements of the SU(1/6) are given by:
$$            g = \exp (i \omega ^I.\beta _I )
\eqno(6)$$
The parameters $\omega $ are real if we use the hermitian generators $\beta
_I$.
There are some rules which will be useful for us later to construct
the SU(1/6)-invariant and covariant quantities:( see$^{26}$ for a general
analysis of $ SU(m/n)$-supergroups and rigorous proofs)\\
i) If $\tilde S$ is a contravariant SU(1/6)-spinor and $U$ is a covariant one
then the product ${\tilde S}.U$ is invariant.\\
ii)The quantity ${\tilde S}\beta _I U$ is a SU(1/6)-covariant vector.\\

The heavy diquark picture for 1HQ-baryons has a very attractive feature that
we can construct an effective theory for heavy diquark by the same route as
HQET.
We are in a lucky situation as the techniques required to do this have been
already worked out by Georgi and Wise$^1$ and by Carone$^2$, who have
constructed effective theories for heavy scalars and for heavy vectors in a
different context.\\

Let us now construct the effective theory for heavy diquarks. The
Foldy-Wouthuysen transformation of heavy quark takes the following form:
$$    Q_v^i(x) = \exp (i M_Q \not v v_\mu x^\mu ).Q^i(x)
\eqno(7)$$
The scalar diquark has the following Foldy-Wouthuysen transformation:\\
$$ {\bar D}_{0a,v}^i(x) = \exp(iM_{D_0}v_\mu x^\mu ). {\bar D}_{0a}^i(x)
\eqno(8)$$
The vector diquark has the following Foldy-Wouthuysen transformation:\\
$$ {\bar D}_{3a,v}^i(x) = \exp(iM_{D_3}v_\mu x^\mu ). {\bar D}_{3a}^i(x)
\eqno(9)$$
So in a supersymmetric form the above Foldy-Wouthuysen transformation can be
written as
follows:
$$     S_v^i(x) =  \exp (-i{\bf M}\not V v_\mu x^\mu ).S^i(x)
\eqno(10)$$
where
$$  \not V ={\bf \Gamma }_\mu. v^\mu          $$
$$ {\bf \Gamma }_\mu = \left (\matrix{
                        \gamma _\mu  & 0     & 0\cr
                        0            & v_\mu & 0\cr
                        0            & 0     & v_\mu \cr}\right )
                        \eqno(11)$$
$$ {\bf M} = \left (\matrix{
                     M_Q   & 0       &0       \cr
                      0    & M_{D_0} &0       \cr
                      0    & 0       &M_{D_3} \cr}\right )
                     \eqno(12)$$
and
$$   v^2= v_\mu .v^\mu = 1
\eqno(13)$$
The matrices $M$ and $\Gamma $ do not act on the SU(1/6)-indices but only
on the spin indices of quark and antidiquarks in a reducible representation.
All terms that refer to the creation and annihilation of heavy particles
are ignored when taking $ M_{Q,D_0,D_3} \longrightarrow \infty $ with the
velocity $ v_\mu $ held fixed, as these processes are suppressed in the limit
we are considering. The leading term in the effective Lagrangian takes the
form :
$$ L_v={i\over 2}({\bar Q}_v v_\mu {\mathord{\buildrel{\lower3pt\hbox
{$\scriptscriptstyle\leftrightarrow$}}\over D}}^\mu Q_v + 2M_D\bar D_{0v}^*
v_\mu {\mathord{\buildrel{\lower3pt\hbox{$\scriptscriptstyle\leftrightarrow$}}
\over D}}^\mu {\bar D}_{0v} - 2M_{D_3} g_{\alpha \beta }{\bar D}_{3v}^{\alpha
*}v_\mu{\mathord{\buildrel{\lower3pt\hbox{$\scriptscriptstyle\leftrightarrow$}}
\over D}}^\mu {\bar D}_{3v}^\beta )
\eqno(14)$$
The color and flavor indices are contracted properly. To write this Lagrangian
in a supersymmetric form let us define
$$ {\bar S}_{v}^i = S_{v}^{i*} \Gamma _0 = S_v^{i*} \left ( \matrix{
           \gamma _0 &0    & 0\cr
           0         &1    & 0 \cr
           0         &0    & -g \cr } \right )
           \eqno(15)$$
where $g$ is the metric matrix. So:

$$ L_v = {i \over 2} (\bar S_v \tilde M v_\mu
{\mathord{\buildrel{\lower3pt\hbox
{$\scriptscriptstyle\leftrightarrow$}}\over D}} ^\mu S_v)
 \eqno(16)$$

where
$$
\tilde M= \left (\matrix{
              1   & 0         &0 \cr
              0   & 2M_{D_0}  &0 \cr
              0   & 0         &2M_{D_3}\cr }\right )
               \eqno(17)$$
We can redefine diquarks so that the $\tilde M$-matrix is unit by the
following normalization
$$  D_a^i    \longleftarrow  { D^i_a \over \sqrt{2M_D} }
\eqno(18)$$
Hence we have an effective Lagrangian
$$
 L_v = {i \over 2} (\bar S_v v_\mu {\mathord{\buildrel{\lower3pt\hbox
{$\scriptscriptstyle\leftrightarrow$}}\over D}} ^\mu S_v)
\eqno(19)$$
which is $SU(1/6)$-invariant manifestly under the following transformations:
 $$ \delta S_v = i\omega _I \beta ^IS_v
 \eqno(20)$$
Let us notice that the supermultiplet is decoupled from the gluon field by
the gauge fixing condition:
$$ v_\mu.A^\mu = 0       \eqno(21)$$
Using the rules mentioned above we have the $SU(1/6)$-covariant vector
currents
$$ J_\mu ^I(1,2) = \bar S_{v1} \tilde {\bf \Gamma }_\mu \beta ^I S_{v2}
\eqno(22)$$
where
$$\tilde {\bf \Gamma }_\mu = \left ( \matrix {
                        \gamma _\mu &0    &0 \cr
                        0      &{\mathord{\buildrel{\lower3pt\hbox{$
         \scriptscriptstyle\leftrightarrow$}}\over \partial }}_\mu  &0 \cr
                        0  &0
&{\mathord{\buildrel{\lower3pt\hbox{$\scriptscriptstyle
         \leftrightarrow$}}\over \partial}}_\mu \cr}
                        \right )
                        \eqno(23)$$
$S_{v1}$ and $S_{v2}$ denote different heavy flavored supermultiplet ( They
can take any of the flavors c,b,t).
The supermultiplet of 1HQ-hadrons can be given by the tensor product of a
light antiquark with the supermultiplet $S$:
$$ H_b = S^i.{\bar q}_{ib} = \left (\matrix{
               Q^i.{\bar q}_{ib} \cr
               {\bar D}_a^i {\bar q}_{ib} \cr}\right )
               \eqno(24)$$
This supermultiplet contains mesons and antibaryons. The light antiquark is
unchanged under the HSUSY transformations of SU(1/6). So $ H_b$ is a
SU(1/6)-covariant spinor.\\
The matrix elements of the above current between the supermultiplet $ H_1$ and
$H_2$ will be $SU(1/6)$-invariant. Therefore by the Wigner-Eckhart theorem all
these matrix elements are determined in terms of a single reduced matrix
element. This is the Isgur-Wise function.\\
For example, if we want to calculate the matrix elements between any two states
$\Psi _{1P}(v')$ and $\Psi _{2Q}(v)$, which are antibaryons or mesons of the
supermultiplets $H_1$ or $H_2$ ( n, l denote the compnents of the hadron
supermultiplets). The matrix elements can now be computed as follows:
$$ \left\langle \Psi _{1P}(v')\right | J_{\mu I} (v',v) \left | \Psi _
{2Q}(v)\right \rangle = -\xi (v'v).Tr ({\bar \Psi }_{1P}\tilde \Gamma _\mu
(v',v) (\beta _I)_P^Q \Psi _{2Q})                            \eqno(25)     $$
 In the expression above we can replace  $i{\mathord{\buildrel{\lower3pt\hbox{$
 \scriptscriptstyle\rightarrow$}}\over \partial _\mu }} $ and $-i{\mathord{
 \buildrel{\lower3pt\hbox{$\scriptscriptstyle\leftarrow$}}\over \partial
 _\mu }}$ by $M_{2}.v_\mu $ and $M_{1}.v'_\mu $ in $\tilde {\bf \Gamma }
 _\mu $ as the heavy supermultiplet is nearly on the mass shell.\\
The heavy supermultiplet satisfies the velocity selection rule:
 $$ \not V.S = S
 \eqno(26)$$
Let us work out the trace formula (26) for the matrix elements of the weak
current $J^{V-A}_\mu $ between the hadron states. To do so we should follow the
steps:\\
i) To give an ansatz for the wavefunctions of the mesons P,$0^-$ and V,$1^+$
and of the baryons
$\Lambda (\Xi )$ , ${1 \over 2}^+$ , $\Sigma (\Omega )$, $ {1 \over 2}^+$ and
$\Sigma ^*
(\Omega ^*)$, $ {3 \over 2}^+$ \\
ii) To substitute the ansatz and the expressions (3)-(5) and (22) for $\beta
_I$
and $\tilde \Gamma _\mu $ into the formula (26) then  work out the trace.
\\
Here we present only the main results:
\\
The ansatz for the mesons'wavefunctions is given as follows:\\
$$   P = \left ( \matrix{
                   \sqrt {M_P}  \gamma _5 (1-\not v )/2 \cr
                                0                     \cr
                                0                    \cr
                          }\right )          \eqno(27) $$
$$   V = \left (\matrix {
                   \sqrt{ M_V}  \not \epsilon (1-\not v)/2 \cr
                                  0                      \cr
                                  0                      \cr
                          } \right )          \eqno(28) $$
where $\epsilon $ denotes the polarization vector of the vector meson $1^+$\\
The ansatz for the baryons'wavefunctions is given as follows:\\
$$ \Lambda = \left ( \matrix{
                        0 \cr
                       u^T C .\sqrt{ 2M_\Lambda} \cr
                        0 \cr
                       }\right )     \eqno (29) $$
where the spinor $ u $ satisfies the velocity selection rule:
$$             \not v u = u    \eqno(30)  $$
and the unconventional normalization:
$$               \bar u u = 1  \eqno(31)  $$
C is the charge conjungation matrix.\\
$$
\Sigma = {1 \over \sqrt{ 6M_\Sigma} } \left (\matrix {
                            0 \cr
                            0 \cr
                       u^T C \sigma^{\mu \nu }v_\nu \gamma _5 \cr
                       } \right )          \eqno(32) $$
$$ \Sigma ^* = {1 \over \sqrt{ 2M_{\Sigma ^*}}} \left (\matrix {
                          0 \cr
                          0 \cr
                          u^{\mu T} C \cr
                        } \right )     \eqno(33) $$
$ u^\mu $ is a Rarita-Schwinger spinor satisfying the equations:
$$   - \bar u_\mu u^\mu  = 1        \eqno(34)  $$
$$ \not v u^\mu  = u^\mu    \eqno(35) $$
We can impose the gauge fixing conditions:
$$ v_\mu u^\mu   = 0       \eqno(36)$$
and
$$ \gamma _\mu u^\mu = 0    \eqno(37)$$
The formula (26) gives the following matrix elements of the weak currents:
$$
< P_{Q_1}(v')| J_\mu ^{V-A} | P_{Q_2}(v)> =  \xi (v'v) (v_\mu +v'_\mu) \sqrt
{M_{P_1}.M_{P_2}}    \eqno(38)$$
$$
  <V_{Q_1}(v')| J_\mu ^{V-A} | V_{Q_2}(v)>
 =  -\xi(v'v)  [ (\epsilon _1^*
\epsilon _2)(v_\mu +v'_\mu) -(\epsilon ^*v)\epsilon_{2 \mu }$$
$$- (\epsilon _2 v)
\epsilon_{1 \mu }]\sqrt{M_{V_1}.M_{V_2}}  \eqno(39) $$
$$
< \Lambda _{Q_1}(v')| J_\mu^{V-A} | \Lambda _{Q_2}(v)> = \xi(v'v){1 \over 2}
(v_\mu +v'_\mu ) \bar u_1 u_2 \sqrt{M_{\Lambda _1}.M_{\Lambda _2}}\eqno(40)$$
$$
 <\Sigma_{Q_1}(v')|J_\mu^{V-A}| \Sigma _{Q_2}(v)>
 =  {1 \over 6} \xi (v'v)[
(2+v'v)(v'_\mu +v_\mu)\bar u_1 u_2 $$
$$ + (1+v'v) \bar u_1 (2\gamma _\mu -v_\mu -v'_\mu )u]\sqrt{M_{\Sigma
_1}.M_{\Sigma _2}} \eqno(41)$$
$$
 <\Sigma^*_{Q_1}(v')|J_\mu ^{V-A}|\Sigma ^*_{Q_2}(v)>
 = {1 \over 2} \xi(v'v) [
-(v'_\mu +v_\mu )\bar u_{1 \mu } u^\mu _2 + \bar u_{1 \mu }v'_\nu u^\nu _2 $$
$$ +
v_\nu \bar u^\nu _1 u_\mu]\sqrt {M_{\Sigma ^*_1}.M_{\Sigma ^*_2}} $$
$$
<\Sigma _{Q_1}(v')|J_\mu ^{V-A}| \Sigma ^*_{Q_2}(v)>
 = {i\over 2\sqrt {3}}
\xi(v'v)[(1+v'v) \bar u_1 \gamma _5 u^\mu _2 $$
$$ + \bar u_1 \gamma _5 (\gamma
_\mu - v_\mu ) v'_\nu u^\nu _2 ]\sqrt {M_{\Sigma _1}.M_{\Sigma ^*_2}}
\eqno(43)$$
We can also compute the light flavour changing matrix elements, which lead to
the same reduced form factor $\xi (v'v)$. The matrix elements between mesonic
and baryonic states vanish because the matrix $\tilde \Gamma _\mu $ is
diagonal.
More details on these matrix elements will be given elsewhere.\\
It is worthy stressing here that new supersymmetry and new model presented
above are not alternative of the LSUSY and the light diquark model. Both of
them
are dynamical symmetries based on certain approximations to baryons. In some
meaning these supersymmetries are similar to the one of Iachello
et.al.$^{27}$
in nuclear physics for heavy nuclei. Specially, in this paper, we have proposed
the heavy diquark model for baryons to go a step beyond the spectator model of
baryons. In this picture one light quark approaches the heavy quark closer than
the other light one and form a heavy diquark. The heavy diquark has a good
chance to survive the gluon exchanges with the remaining light quark. So, the
last
could be approximated very well as a
spectator. We have constructed a supersymmetric effective model for heavy
quark and heavy diquark. In this supersymmetric model the matrix elements of
weak currents are characterized by one universal Isgur-Wise function and the
weak formfactors of baryons collapse to the one of mesons
 \thanks{thanks are due to F.Hussain who remind us the importance of
 this consequence.}. We can also combine the heavy quark spin symmetry
with the SU(1/6) supersymmetry and put all spin components of the heavy quark
and heavy antidiquark together into a 14-dimensional spinor of the supergroup
SU(2/12). Combining with the heavy flavour symmetry $SU(3)$ of Isgur and Wise
the total symmetry of the "heavy world" is indeed tremendous with the
supersymmetry $SU(6/36)$. This is the largest symmetry we know about this
world. So we come to the conclusion that the set of 1HQ-hadrons is governed
by a simpler law than we thought.   \\

{\bf  Acknowledgements }\\

  I would like to thank Professor F.Hussain for helpful discussions and
hospitality at
International Centre for Theoretical Physics (Trieste), where this work has
been done.

  I am also grateful to Professor Kameshwar C.Wali for some corrections.\\

\end{document}